\documentclass[12pt]{article}
\usepackage{graphicx}
\usepackage{feynmf} 
\begin{document}
\begin{fmffile}{fmfd1} 

\thispagestyle{empty}
\mbox{ }
\rightline{April 2000}\\
\vspace{2.0cm}
\begin{center}
\begin{Large}
{\bf Thermal Renormalons in Scalar Field Theory\\}
\end{Large}
\vspace{.7cm}
{\bf  M. Loewe   and C. Valenzuela} \\
\vspace{.7cm}

Facultad de F\'{\i}sica, Pontificia Universidad Cat\'olica
de Chile\\
Casilla 306, Santiago 22, Chile\\
\end{center}
\vspace{.5cm}


\begin{abstract}
\noindent
In the frame of the scalar theory $g \phi ^{4}$, we explore
the occurrence of thermal renormalons, i. e. 
temperature dependent singularities in the Borel plane.
The discussion of a particular renormalon type diagram at finite temperature,
using Thermofield Dynamics,
allows us to establish that these singularities actually get a temperature dependence.
This dependence appears in the residues of the poles,
remaining their positions unchanged with temperature.
\end{abstract}
\newpage


\section{Introduction}

\bigskip 
\indent 
\indent During the last years, there have been an impressive 
amount of theoretical work
on renormalons in different scenarios. For a recent review see \cite{Beneke}. 
One of the
main motivations behind this effort concerns the non perturbative
structure of Quantum Chromodynamics (QCD). On the other side, 
finite temperature effects have also called the attention of many authors \cite{cad},
due to their crucial role in understanding thermal aspects of
the hadron dynamics, whith  special emphasis on the deconfining phase transition
and the production of the quark gluon plasma \cite{rev}. 
It seems natural, therefore, to start a systematic study of the occurrence and also
the possible phenomenological role of thermal renormalons. Here we discuss, as  a
first step in this direction, the scalar theory $g \phi ^{4}$. In spite of being
a non realistic approach for phenomenological purposes, this analysis will allow us to gain
a first impression on this kind of effects.

\medskip
The fundamental characteristic of a renormalon type diagram 
is an insertion of at least a chain of bubbles in a loop diagram,
and behaves like $k!$ for large values of $k$, being $k$ the number of bubbles in the chain.
These diagrams, in the usual zero temperature situation, induce
the existence of certain types of poles in the Borel plane. 
In this paper we explore in detail the extension to the finite temperature scenario
of one  kind 
of renormalon type Feynman diagrams, see Fig. 1, 
that contribute to the two-point function. The set of diagrams we have chosen
is an example of a renormalon type contribution (ultraviolet renormalons, UVR), and gives us
a hint about the properties of the thermal Borel plane, i.e. positions and residues of poles 
as a function of temperature.

\medskip
The discussion at finite temperature has been done
using the machinery of Thermo Field Dynamics (TFD) \cite{TFD}.
The main result from our analysis is 
that residues can get an explicit dependence on temperature, 
remaining, nevertheless, the
position of the poles in the Borel plane unchanged.


\medskip
The plan of this paper is as follows. In section 2,
the analysis at zero temperature of a particular set of Feynman diagrams, see Fig. 1, 
that contribute to the two-point function in the $g \phi ^{4}$ theory,
allows us to show the presence of a renormalon type singularity in the Borel plane. 
Section 3 is devoted to 
a brief discussion of how to handle the chain 
of bubbles shown in Fig. 2 at finite temperature.
This will be done in the deep euclidean region,
i.e. $-p^{2}\gg m^{2}$, of the momentum $p$ that circulates through the chain of bubbles.
Using these results, in section 4 we calculate  
our renormalon type diagram (Fig. 1) at finite temperature
and proceed
to present our conclusions. 


\section{The zero temperature renormalon}

\bigskip
\indent 
\indent
In the present paper the  diagram shown in Fig. 1  will be taken as a source
for  renormalons. Our goal is to explore the influence of temperature on this type
of diagrams. Let us call by
$R_k(p)$ the diagram of order $k$, where $k$ denotes the number of vertices. In order to
establish our notation, we will first 
review the zero temperature calculation for the  renormalon asociated
to this diagram \cite{COLLECT}.


\begin{figure}[h]
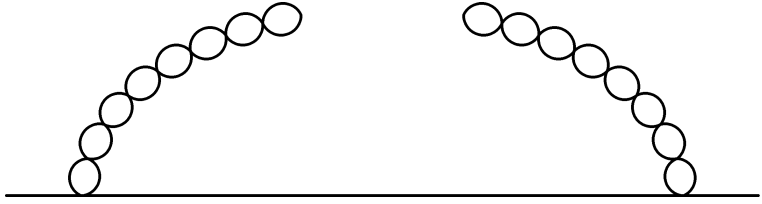

\begin{center}
\begin{fmfchar*}(100,25)

 \fmfstraight     \fmfleftn{i}{2}     \fmfrightn{o}{2}

 \fmfforce{.1w,0}{a1}           \fmfforce{.9w,0}{a21}
 \fmfforce{.108w,.198h}{a2}     \fmfforce{.892w,.198h}{a20}
 \fmfforce{.13w,.38h}{a3}       \fmfforce{.87w,.38h}{a19}
 \fmfforce{.162w,.534h}{a4}     \fmfforce{.838w,.534h}{a18}
 \fmfforce{.201w,.664h}{a5}     \fmfforce{.799w,.664h}{a17}
 \fmfforce{.245w,.77h}{a6}      \fmfforce{.755w,.77h}{a16}
 \fmfforce{.292w,.854h}{a7}     \fmfforce{.708w,.854h}{a15}
 \fmfforce{.341w,.918h}{a8}     \fmfforce{.659w,.918h}{a14}
 \fmfforce{.392w,.962h}{a9}     \fmfforce{.608w,.962h}{a13}
 \fmfforce{.445w,.99h}{a10}     \fmfforce{.555w,.99h}{a12}
 \fmfforce{.5w,h}{a11}
 
 \fmf{plain}{i1,a1} \fmf{plain,tension=.15}{a1,a21} \fmf{plain}{a21,o1}
 
 \fmf{plain,left=0.8,tension=.1}{a1,a2,a1}
 \fmf{plain,left=0.8,tension=.1}{a2,a3,a2}
 \fmf{plain,left=0.8,tension=.1}{a3,a4,a3}
 \fmf{plain,left=0.8,tension=.1}{a4,a5,a4}
 \fmf{plain,left=0.8,tension=.1}{a5,a6,a5}
 \fmf{plain,left=0.8,tension=.1}{a6,a7,a6}
 \fmf{plain,left=0.8,tension=.1}{a7,a8,a7}
 \fmf{plain,left=0.8,tension=.1}{a8,a9,a8}
 \fmf{plain,left=0.8,tension=.1}{a13,a14,a13}
 \fmf{plain,left=0.8,tension=.1}{a14,a15,a14}
 \fmf{plain,left=0.8,tension=.1}{a15,a16,a15}
 \fmf{plain,left=0.8,tension=.1}{a16,a17,a16}
 \fmf{plain,left=0.8,tension=.1}{a17,a18,a17}
 \fmf{plain,left=0.8,tension=.1}{a18,a19,a18}
 \fmf{plain,left=0.8,tension=.1}{a19,a20,a19}
 \fmf{plain,left=0.8,tension=.1}{a20,a21,a20}
 
 \fmfv{l=$......$,l.a=0}{a10}
                                
\end{fmfchar*}
\end{center}

\caption{Renormalon type contribution to the two-point function.}
\end{figure}


If we denote by $B(l)$ the one loop correction to the four point function, the so called 
``fish" diagram \cite{ram}, we have

\begin{equation}
R_k(p)=\int\frac{d^{4}l}{(2\pi)^{4}} \frac{i}{(p+l)^{2}-m^{2}+i\epsilon}
       \hspace{0.1cm}\frac{1}{(-ig)^{k-2}}[B(l)]^{k-1},
\end{equation}

\noindent
where the $(-ig)^{k-2}$ factor is due to the double counting of vertices in $B(l)$.
The relevant contribution to the integral comes from the deep euclidean region.
Therefore, it is enough to approximate $B(l)$ in this region.

\begin{displaymath}
B(l) = \frac{(-ig)^{2}}{2} \int\frac{d^{4}q}{(2\pi)^{4}} 
       \frac{i}{q^{2}-m^{2}+i\epsilon} \,\frac{i}{(q+l)^{2}-m^{2}+i\epsilon}.
\end{displaymath}

\begin{equation}
       \stackrel{-l^2\rightarrow \infty}{\longrightarrow}  
       \qquad\frac{-ig^{2}\quad}{32\pi^{2}}\log(-l^{2}).\qquad\qquad\qquad\qquad
\end{equation}

\noindent
(the argument of the logarithm is in mass units)

In this way we have
\begin{equation}
R_k(p)=\frac{-ig^{k}\quad}{(32\pi^2)^{k-1}}
       \int\frac{d^{4}l}{(2\pi)^{4}} \frac{1}{(p+l)^{2}+m^{2}}
       (\log(l^{2}))^{k-1},
\end{equation}

\noindent
where a Wick rotation has been performed.

\medskip
This expression is ultraviolet divergent. 
Since $g \phi^4$ is a renormalizable theory,
we can concentrate on the finite part.
In order to do this the propagator 
is expanded in powers of $1/l^2$. The first two ultraviolet divergent 
terms are subtracted,
and we keep only the first
convergent term in the expansion. This procedure, which at the end induces one pole
in the Borel plane, i.e. one renormalon, will be followed also in the next sections,
where the finite temperature
corrections will be computed. Due to this expansion,
the dependence on $p$, the external momentum of the two point function, actually dissapears. So we find

\begin{eqnarray}
R_k \propto -i\bigg(\frac{g}{32\pi^{2}}\bigg)^{k}        \int dl
                   \frac{1}{l^3}(\log(l^{2}))^{k-1}.
\end{eqnarray}

In the last expression we can see that the main contribution 
to $R_k$, at large values of $k$,  comes 
from large values of $l$, that corresponds to the deep euclidean region, 
since we have done a Wick rotation. 
Introducing the new variable
$l=e^t$, we can see that $R_k$ becomes proportional to the gamma function.

\begin{equation}
R_k\propto -i\bigg(\frac{g}{32\pi^{2}}\bigg)^{k}\Gamma(k).
\end{equation}

Let us remain breifly the idea of the Borel transform.
If, in general, we have a divergent series in terms of
a certain expansion parameter $a$, of the form:

\begin{equation}
D[a]=\sum_{n=1}^\infty D_na^n,       
\end{equation}

\noindent
then, one possibility to give a meaning to this series 
is to make use of the Borel transform method \cite{hardy},
according to which a new perturbative expansion, in a new expansion parameter b,
 is considered
by dividing each coefficient of the previous series by $n!$ in the following way

\begin{equation}
B[b] = \sum_{n=0}^\infty D_{n+1}\frac{b^n}{n!}.   
\end{equation}

Formally, we see that the $D[a]$ corresponds to the Laplace
transform of $B[b]$.
Using this idea in our case, we can introduce $B[b]$ as the Borel transform of $R_k$
according to

\begin{eqnarray}
B[b]  &=&  \sum \Bigg(\frac{R_k}{g^k}\Bigg)
                \frac{b^{k-1}}{(k-1)!},     \nonumber   \\
      &\propto & \frac{-i}{1-b/{32\pi^2}}. 
\end{eqnarray}
				
The $k!$ behavior of $R_k$, for large values of $k$,
is responsible for the appearence of
a pole in the real axis of the Borel plane, 
i.e. in the integration range of the Laplace transform.
The meaning of this or other poles in the Borel plane
corresponds to an ambiguity in the resummation of the series.
Physically, this imply the existence of essential errors 
associated to perturbatives expansions of physical magnitudes.


\section{Chain of bubbles at finite temperature}

\bigskip
\indent 
\indent
In this section we revise the finite temperature 
calculation for the chain of bubbles shown in Fig. 2, which
contribute to the four-point function. The original
discussion can be found in \cite{Fuji}. Later we will approximate
our result for deep euclidean values of the momentum that goes through the chain.
Let us denote by $I_k$ the sum of all diagrams of the form shown in Fig. 2,
with $k$ bubbles and fixed external vertices of type one, according to the rules of TFD,
including a global imaginary factor $i$. The sum is over all
possible combinations of internal type of vertices,  
which, as we know, can be of first or second type.


\begin{figure}
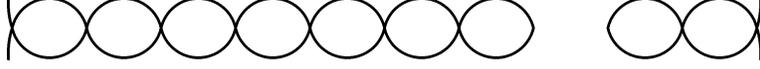

\begin{center}
\begin{fmfchar*}(100,25)

  \fmfleft{f1,i1,i2,f2}    \fmfright{f3,o1,o2,f4}            
  \fmf{plain,left=.1,tension=2}{i1,v1,i2}
  \fmf{plain,left=0.8,tension=.1}{v1,v2,v1}
  \fmf{plain,left=0.8,tension=.1}{v2,v3,v2}
  \fmf{plain,left=0.8,tension=.1}{v3,v4,v3}
  \fmf{plain,left=0.8,tension=.1}{v4,v5,v4}
  \fmf{plain,left=0.8,tension=.1}{v5,v6,v5}
  \fmf{plain,left=0.8,tension=.1}{v6,v7,v6}
  \fmf{plain,left=0.8,tension=.1}{v7,v8,v7}
  \fmf{phantom,left=0.8,tension=.1}{v8,v9,v8}  
      \fmfv{l=$...$,l.a=0,l.d=.028w}{v8}
  \fmf{plain,left=0.8,tension=.1}{v9,v10,v9}
  \fmf{plain,left=0.8,tension=.1}{v10,v11,v10}
  \fmf{plain,right=.1,tension=2}{o1,v11,o2}

\end{fmfchar*}
\end{center}

\caption{The chain of bubbles present in Fig. 1.}
\end{figure}


\medskip
As it was shown in \cite{Fuji}, and after correcting some missprints, 
$I_k$ can be expressed as a function of $I_1$ according to

\begin{eqnarray}
  Re\, I_k &=& \frac{g}{2}(\alpha^k+\gamma^k).            \\
  Im\, I_k &=& \frac{-ig\quad}{2}  
             \bigg(\frac{e^{\beta \arrowvert p_0\arrowvert}+1}
                        {e^{\beta \arrowvert p_0\arrowvert}-1}\bigg)
             (\alpha^k-\gamma^k).
\end{eqnarray}

\noindent 
where

\begin{equation}
\alpha(\gamma) = \frac{A\pm iB}{g}.
\end{equation}

\noindent
In the previous expression $\alpha $ corresponds to the plus and $\gamma $
to the minus sign, respectively, and $p$ is the momentum that circulates through the chain. 
$A$ and $B$ are given by

\begin{eqnarray}
A &=& Re\,I_1,                                     \\
B &=& \frac{e^{\beta \arrowvert p_0\arrowvert}-1}
         {e^{\beta \arrowvert p_0\arrowvert}+1} \quad Im\,I_1,
\end{eqnarray}

\noindent
So, for $I_1$, the ``fish" diagram at finite temperature including the global
factor $i$, we have

\begin{eqnarray}
I_1 = i(-ig)^2\frac{1}{2} \int\frac{d^{4}l}{(2\pi)^4} 
       \Bigg\{\frac{i}{l^2-m^2+i\epsilon}+
        \frac{2\pi\delta(l^2-m^2)}{e^{\beta | l_0 |}-1} \Bigg\}
       &     \nonumber \\ 
  \times\Bigg\{\frac{i}{(l+p)^2-m^2+i\epsilon}+
        \frac{2\pi\delta((l+p)^2-m^2)}
                             {e^{\beta \arrowvert l_0+p_0\arrowvert}-1} \Bigg\}&.
\end{eqnarray}

It is convenient at this point to give
the explicit expressions for the real and imaginary parts of 
$I_{1_0}$ and $I_{1_\beta}$, zero and finite temperature parts of $I_1$, respectively,
such that $I_1=I_{1_0}+I_{1_\beta}$.

\begin{eqnarray}
Re\,I_{1_0}           &=& 
   \frac{g^2}{32\pi^2}
   \sqrt{\frac{|4m^2-p^2|}{|p^2|}} 
   \log\Bigg( \frac{\sqrt{|4m^2-p^2|}+\sqrt{|p^2|}}
                   {\sqrt{|4m^2-p^2|}-\sqrt{|p^2|}}  
                   \Bigg).     \label{reaal}  \\  
Im\,I_{1_0}           &=&    
   \theta(p^2-4m^2)\,\frac{(-g^2)}{32\pi}\sqrt{1-\frac{4m^2}{p^2}}. \\                        
Re\,I_{1_\beta}       &=&
   \frac{g^2}{16\pi|\vec{p}|} 
   \int_{0}^{\infty}\frac{dl\,l}{E(e^{\beta E}-1)}
   \log\Bigg(\Bigg|\frac
   {(2p_0 E)^2-(2l|\vec{p}|+p^2)^2}
   {(2p_0 E)^2-(2l|\vec{p}|-p^2)^2}
   \Bigg|\Bigg).       \label{eq15}    \\
Im\,I_{1_\beta} &=& \frac{-g^2}{8}\int\frac{d^3l}{(2\pi)^2 E_1 E_2}
   \Bigg(\delta(p_0+E_1+E_2)+\delta(p_0-E_1-E_2)   \nonumber    \\   
& & \qquad\qquad\qquad\quad\:\:
    +\delta(p_0+E_1-E_2)+\delta(p_0-E_1+E_2) \Bigg)   \nonumber     \\
& & \qquad\quad\times\Bigg(\frac{1}{e^{\beta E_1}-1}+\frac{1}{e^{\beta E_2}-1}+
    \frac{2}{(e^{\beta E_1}-1)(e^{\beta E_2}-1)}   \Bigg). \label{fourdelta}
\end{eqnarray}

In the previous formulae
$E_{1} = \sqrt{\vec{l}^{2} + m^{2}}$ and 
$E_{2}= \sqrt{(\vec{l} + \vec{p})^{2} + m^{2}}$,
and the imaginary part has been obtained following
 \cite{Kobes}. 
Since $\alpha$ and $\gamma$ are conjugated to each other, 
we can rewrite

\begin{eqnarray}
\label{rik}
  Re\, I_k &=& ga^k\cos(k\theta),          \\
\label{iik}
  Im\, I_k &=& ga^k\sin(k\theta)  
             \bigg(\frac{e^{\beta \arrowvert p_0\arrowvert}+1}
                        {e^{\beta \arrowvert p_0\arrowvert}-1}\bigg),
\end{eqnarray}

\noindent
where we have introduced
\begin{eqnarray}
  a      &=& \frac{\sqrt{A^2+B^2}}{g}.      \label{defa}      \\
  \theta &=& \arctan\bigg(\frac{B}{A}\bigg). \label{theta}
\end{eqnarray}

From the zero temperature calculation, we know that the renormalon type
contribution of the diagram shown in Fig. 1 comes from the deep euclidean 
region, specifically for $\sqrt{-p^{2}}\approx e^{k/2}$,
where $k$ is the number of bubbles in the chain, 
and $p$ the momentum that circulates through the chain.
In what follows,
we will examine the chain of bubbles in the deep euclidean 
region at  finite temperature,  assuming $\sqrt{-p^{2}} \approx e^{k/2}$.
Later, by replacing 
the explicit expression for our chain of bubbles
shown in Fig. 1,  we will check if 
the previous condition,  
$\sqrt{-p^{2}}\approx e^{k/2}$, is actually realized.

\medskip
First we calculate $Im \,I_{1_\beta}$ and $ Re\, I_{1_\beta}$ for $T\neq 0$.
From the four $\delta $'s that appear
in $Im \, I_{1_{\beta }}$, eq. \ref{fourdelta}, it is easy to see that only those whose arguments include
energy differences
survive in the deep euclidean limit. So we have

\begin{equation}
Im\,I_{1_\beta} \approx \frac{-g^2}{8}\int\frac{d^3l}{(2\pi)^2 E_1 E_2}
   2\delta(E_1-E_2)
   \Bigg(\frac{2}{e^{\beta E_1}-1}+\frac{2}{(e^{\beta E_1}-1)^2}   
   \Bigg).
\end{equation}

By integrating and considering that $e^{\beta \mid \vec{p} \mid /2} \gg 1$,
we have finally 

\begin{equation}
Im\,I_{1_\beta} \approx \frac{-g^2}{4\pi}
\frac{e^{-\beta |\vec{p}|/2}}{\beta |\vec{p}|}.
\end{equation}

Turning to the real part of $I_{1_\beta}$, eq. \ref{eq15}, note that the argument of the 
logarithm in the deep euclidean limit can be approximated in such a way that
the integral can be written as

\begin{equation}
Re\,I_{1_\beta} \approx
   \frac{g^2}{16\pi|\vec{p}|} 
   \int_{0}^{q} \frac{dl\,l}{E(e^{\beta E}-1)}
   \,\, 2\log\Bigg(1-\frac{4l|\vec{p}|}{(-p^2)}\Bigg),
\end{equation}

\noindent
where $-p^{2}$ is sufficiently large so that we can reach
the main contribution to the
integral in the region where the logarithm can be expanded in powers
of $4l|\vec{p}|/(-p^{2})$. In the previous expression, $q$ denotes
a certain bound for the integration in $l$ such that for  values of $l$ bigger than $q$,
the contribuition to the integral turns out
to be negligible due to the exponential supression. 
At the end we can take $q\rightarrow \infty$. The real part, then, can be written as a
series in powers of $1/\sqrt{-p^{2}}$

\begin{equation}
\label{real2}
Re\,I_{1_\beta} \approx \frac{g^2}{32\pi^2}
  \sum_{n=2}^\infty
  f_n(\beta)\Bigg(\frac{\: 1}{-p^2}\Bigg)^{n/2},
\end{equation}

\noindent
where the coefficients, which in our limit turn out to be essentially
independent of the external momentum,  
 are given by

\begin{equation}
f_n(\beta) \approx -4\pi \int_{0}^{\infty}
    \frac{dl\,l}{E(e^{\beta E}-1)}
    \frac{\quad(4k)^{n-1}}{n-1}.
\end{equation}

Using the previous results in eq. \ref{theta}, we can see that

\begin{equation}
|k\theta |\ll 1,
\end{equation}

\noindent
where, once again, we have taken the assumption $\sqrt{-p^{2}}\approx e^{k/2}$.
Using this fact, we can approximate in eqs. \ref{rik}  and \ref{iik}
$\cos(k\theta) \approx 1$ and $\sin(k\theta)\approx 0$.
This means that $I_{k} \approx ga^{k}$. 
Since $B$ in eq. \ref{defa}
can be neglected then $I_{k}\approx g(A/g)^{k}$. 
Therefore, using our
expressions for the real part of $I_{1}$, eqs. \ref{reaal} and \ref{real2}, 
finally we get

\begin{equation}
\label{app}
I_k \approx g \Big(\frac{g}{32\pi^2}\Big)^k \Bigg[
  (\log(-p^2))^k + k (\log(-p^2))^{k-1} 
  \frac{f_2(\beta)}{-p^2} +\cdot\cdot\cdot \Bigg].
\end{equation}

This is the fundamental expression that we will use in the next section
for the discussion of our diagram at finite temperature.
Note that the leading or first term, which does not depend
on temperature, is the same one we found in the zero temperature analysis. 
The second term is the first thermal contribution. The dots denote higher order 
terms in the expansion in powers of $1/p^2$.

It is interesting to mention that we can find an approximated 
expression for the coefficients $f_{n}(\beta )$ in the low 
temperature region, where
$\beta m \gg 1$ 

\begin{equation}
f_n(\beta) \approx \frac{-4^n\pi}{2\,(n-1)}  \int_0^{\infty}
                   \frac{dx\: x^{(n-1)/2}}{\sqrt{x+1}}\:
                   e^{-\beta\sqrt{x+1}},
\end{equation}

\noindent
which can be evaluated exactly, see \cite{Grad}

\begin{equation}
f_n(\beta) \approx \frac{-4^n\sqrt\pi}{n-1}  \Bigg(\frac{2}{\beta}\Bigg)^{n/2}
                   \Gamma(\frac{n+1}{2})\:
                   K_{-n/2}(\beta),
\end{equation}

\noindent
and where $K_{\nu }$ are Bessel type functions.


\section{Renormalons at finite temperature}

\bigskip
\indent 
\indent
Using  eq. \ref{app} for the chain of bubbles,
the renormalon diagram shown in Fig. 1
will be calculated  in the frame of TFD.
At finite temperature, $R_k(T)$ denotes the sum of all diagrams of the shape 
shown in Fig. 1 with $k$ vertices, being the external
vertices of the first type. As it was the case in the chain of bubbles,
the sum is over all possible combinations 
of internal type of vertices. In order to use our expression for $I_k$, 
this sum must be performed before the integral over the internal momentum
that circulates through the chain of bubbles.
Here we will obtain an expression for $R_k(T)$, finding 
the location of the induced poles in the Borel plane and the corresponding residues.

\medskip
The expression we have to deal with is

\begin{eqnarray}
\label{c}
R_k (p) = \int\frac{d^{4}l}{(2\pi)^4} 
            \Bigg(\frac{i}{(p+l)^2-m^2+i\epsilon}+
                   \frac{2\pi\delta((p+l)^2-m^2)}
                        {e^{\beta |p_0+l_0|}-1} \Bigg)
            \frac{I_{k-1}(l)}{i}.                          
\end{eqnarray}

The zero temperature contribution arises from the zero temperature
part of the propagator times the zero temperature part of $I_{k-1}$.
We would like to discuss
the following two cases: a) the product of the thermal
part of the propagator times 
the zero temperature part of $I_{k-1}$, denoted by $R_{k}^{a}$,
and b) the product of  the usual zero temperature propagator
times $I_{k-1}$ (including the zero temperature part of $I_{k-1}$), denoted by $R_{k}^{b}$.

\medskip
Let us start with case a).  We have

\begin{equation}
R_{k}^{a}(p) = \int\frac{d^{4}l}{(2\pi)^4} 
\Bigg(\frac{2\pi\delta((p+l)^2-m^2)}
           {e^{\beta |p_0+l_0|}-1}\Bigg)
\Bigg(\frac{-ig^k}{(32\pi^2)^{k-1}} (\log (-l^2))^{k-1}
\Bigg).
\end{equation}

The delta function excludes the deep euclidean region in the momentum $l$.
In spite of this fact, we can exime the limit $|\vec{l}| \gg m$, after the integration in $l^{0}$
has been done.
It turns out that $R_{k}^{a}(p)$ is proportional to

\begin{equation}
R_{k}^{a}(p) \propto -i \Bigg(\frac{g}{32\pi^2}\Bigg)^k    \int dl \:
\frac{l}{e^{\beta l}-1} (\log l)^{k-1}.
\end{equation}

The sum over $k$ in the expression above is Borel summable and, therefore, it
does not imply any renormalon.

\medskip
The second case, case b), corresponds to

\begin{eqnarray}
\label{b}
\lefteqn{R_{k}^b(p) = \int\frac{d^{4}l}{(2\pi)^4} 
\Bigg(\frac{1}{(p+l)^2-m^2+i\epsilon}\Bigg)        
\frac{g^k}{(32\pi^2)^{k-1}}}  \qquad \qquad        \nonumber         \\ 
& & \times\Bigg[
          (\log(-l^2))^{k-1} + (k-1) (\log(-l^2))^{k-2} 
          \frac{f_2(\beta)}{-l^2}+\cdot\cdot\cdot  \Bigg].                      
\end{eqnarray}

After the Wick rotation and substracting the divergents terms, exactly
in the way as we did 
in section 2,
we find that the last
expression is proportional to

\begin{equation}
\label{a}
R_{k}^b \propto -i\Bigg(\frac{g}{32\pi^2}\Bigg)^k \int dt \: e^{-2t} \Bigg[
(2t)^{k-1}+(k-1)(2t)^{k-2}e^{-2t}f_2(\beta)   +\cdot\cdot\cdot       \Bigg],                        
\end{equation}

\noindent
where the maxima of the first and  second terms in the integral are reached 
at $t = \log(\sqrt{-l^{2}})$ $\approx k/2$ and $t \approx k/4$, respectively. Note
that the assumption we made in the previous section, 
$\sqrt{-p^{2}}\approx e^{k/2}$, has changed now in
the second term, due to the factor 2 in the exponential. 
This fact does not affect our approximations.

\medskip
Doing the integral, we see that $R_{k}^b$ is proportional to

\begin{equation}
\label{gral}
R_{k}^b \propto -i\Bigg(\frac{g}{32\pi^2}\Bigg)^k\Gamma (k)
-2\, i\, f_2(\beta)\Bigg(\frac{g}{64\pi^2}\Bigg)^k\Gamma (k)+\cdot\cdot\cdot.   
\end{equation}

We see that those terms proportional to the product of the temperature
dependent part of the propagator times
the whole series of $I_{k}$ are Borel summable, since
the leading term in the series of $I_{k}$, case a), is already Borel summable.
On the contrary, the zero temperature part of the propagator
times $I_{k}$ gives us a series of terms that behave like $k!$
being, therefore, non Borel summable.  We have calculated explicitly the first 
temperature dependent term of this series associated to a $k!$ behavior.

\medskip
The Borel transform, $B[b]$, of the sum $\sum R_{k}$, taking
into account only those non-summable Borel terms, is given by:
 
 \begin{eqnarray}
B[b]  &=&  \sum \Bigg(\frac{R_k}{g^k}\Bigg)
                \frac{b^{k-1}}{(k-1)!},     \nonumber   \\
      &\propto & -i\Bigg[\frac{1}{1-b/{32\pi^2}} + 2\,f_2(\beta)\,
                         \frac{1}{1-b/{64\pi^2}} +\cdot\cdot\cdot
                 \Bigg]. 
\end{eqnarray}

The first term in the last expression is the zero temperature renormalon we found 
in section 2.
The second term  corresponds to the first thermal 
singularity in the Borel plane, and the dots
denote a whole series of poles. 
The structure of singularities in the Borel plane, associated to our 
renormalon type diagram, is shown in Fig. 3.
The leading or first pole along the real axis is the zero temperature renormalon.
The series of new thermal poles, i.e. thermal renormalons, 
to the right side of the first one, are subleading contributions
to the Borel sum ambiguity.


\begin{figure}
\begin{center}
\includegraphics[scale=.7]{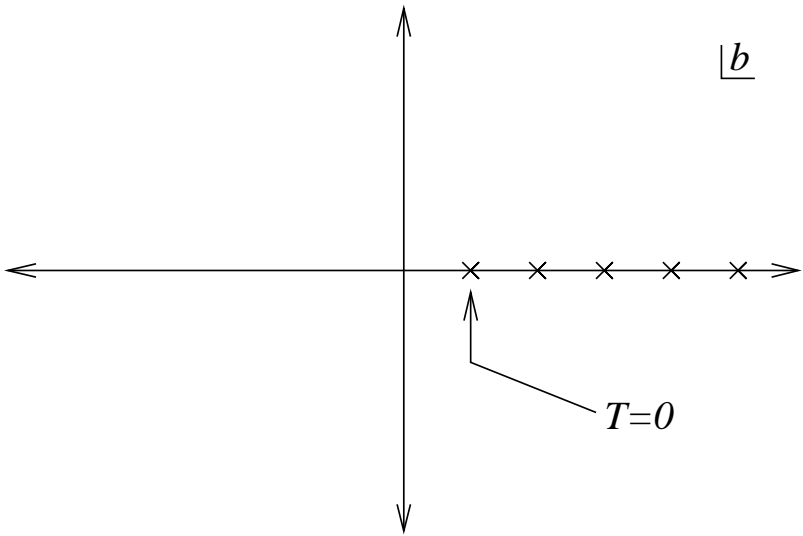}
\end{center}
\caption{Singularities in the Borel plane.}
\end{figure}


Summarizing, in this paper we have discussed thermal corrections to a particular
renormalon type diagram in the theory $g \phi^4$. From this analysis, we show
the existence of thermal singularities in the Borel plane associated to this particular
diagram. The main properties of these thermal renormalons are the following:
a) Their location in the Borel plane does not depend on temperature.
They are situated at the  points where the zero temperature renormalons 
are usually located, i.e. at
$n/\beta_0$, being $n$ an integer and 
$\beta_0$ the first coefficient of the $\beta$ function \cite{alt}.
b) Their residues, on the contrary,
have an explicit dependence on temperature, through the factors
$f_{n}(\beta )$ that vanish when $T \rightarrow 0$. 

The conclusions from the particular diagram we have discussed here,
suggest us to conjecture that these will be actually 
general properties of the Borel plane associated to
correlation functions at finite temperature: 
residues, in general, will depend on temperature 
being, however, the location of the poles temperature independent.
Thermal renormalons in the scalar theory are always subleading
in the ambiguity of Borel sum, 
i.e. the leading pole is temperature independent.
This fact is related to the ultraviolet character of the
renormalons in this theory. Thermal corrections are associated
to long distance correlations in the system, whereas the leading renormalons
come from the singular behavior at short distances.
Something equivalent happens with the axial anomaly, where it has been shown
that the anomaly itself does not depend on temperature \cite{lc}.
In QCD this situation will be probably different 
due to the existence of infrared renormalons.


\section*{Acknowledgments}
This work has been supported by FONDECYT under contract No. 1980577. C. V. acknowledges
the support of a Magister fellowship by the P. Universidad Cat\'olica de Chile. The
authors would like to thank A. Actor for helpful discussions.



\end{fmffile}
\end{document}